# ENHANCING QoS AND QoE IN IMS ENABLED NEXT GENERATION NETWORKS

Kamaljit I. Lakhtaria

Atmiya Institute of Technology & Science,
Rajkot, Gujarat, INDIA
Email: kamaljit.ilakhtaria@gmail.com

*ABSTRACT*

*Managing network complexity, accommodating greater numbers of subscribers, improving coverage to support data services (e.g. email, video, and music downloads), keeping up to speed with fast-changing technology, and driving maximum value from existing networks – all while reducing CapEX and OpEX and ensuring Quality of Service (QoS) for the network and Quality of Experience (QoE) for the user. These are just some of the pressing business issues faced by mobileservice providers, summarized by the demand to "achieve more, for less." The ultimate goal of optimization techniques at the network and application layer is to ensure End-user perceived QoS. The next generation networks (NGN), a composite environment of proven telecommunications and Internet-oriented mechanisms have become generally recognized as the telecommunications environment of the future. However, the nature of the NGN environment presents several complex issues regarding quality assurance that have not existed in the legacy environments (e.g., multi-network, multi-vendor, and multi-operator IP-based telecommunications environment, distributed intelligence, third-party provisioning, fixed-wireless and mobile access, etc.). In this Research Paper, a service aware policy-based approach to NGN quality assurance is presented, taking into account both perceptual quality of experience and technology-dependant quality of service issues. The respective procedures, entities, mechanisms, and profiles are discussed. The purpose of the presented approach is in research, development, and discussion of pursuing the end-to-end controllability of the quality of the multimedia NGN-based communications in an environment that is best effort in its nature and promotes end user's access agnosticism, service agility, and global mobility.*

## 1.   INTRODUCTION

The communications are no longer limited to the choice of voice, data, or video: their multimedia nature presumes an enhanced end user's experience engaging various services and contents within a single convergent session. Commonly understood as the next generation networks (NGN), a composite environment of proven telecommunications and Internet-oriented mechanisms is established, enabling agile service creation, access agnosticism, and global mobility of end users. The NGN environment is based on the Internet protocol (IP) transport platform and adopts a model of a transparently separated service provisioning platform above a heterogeneous transport and access platform, employing various technologies to accomplish the IP connectivity. Unlike legacy solutions, the NGN tends to be access agnostic; from the functional viewpoint it consists of subsystems—logical groupings of entities that perform precisely defined functionalities— which originate from both fixed and wireless domains and promote unlimited choice of access possibilities (e.g., fixed—DSL, cable—or wireless —UMTS, WiMAX, WiFi). The key objective of the NGN environment is to converge and turn to advantage the benefits of the two communications worlds by combining the controllability, reliability, and quality of telecom with the flexibility, ease of operation, creativeness, and end users' involvement of the Internet.

*A. Quality Mechanisms*

The measure of system performance represents one of the basic evaluation criteria of a successful network, solution or a service from nearly all viewpoints: deployment, operation, and customer satisfaction.

In general referred to as the quality, there are basically two approaches to defining, measuring and assessing the success of meeting a specific set of requirements or an expected behavior. The measure of performance from the network perspective is known as the quality of service (QoS) and involves a range of QoS mechanisms that are implemented for the purpose of meeting the defined conditions in the network. Typically, QoS metrics include network operation parameters (i.e., bandwidth, packet loss, delay, and jitter). On the other hand, the measure of performance as perceived from the end user is known as the quality of experience (QoE) and addresses the overall satisfaction of the end user and the ability to meet their expectations. While the QoS is rather objective approach to assessing the success of performing within a specified network subsection, the QoE is subjective, measured on an end-to-end basis, and involves human-related criteria, based on which certain descriptive indexes of performance are set. Some examples of QoE metrics are the mean opinion score (MOS), degraded seconds, errored seconds, unavailable seconds, etc.

When the network, service, or solution engineering is discussed from the quality viewpoint, there are generally two approaches available:

- The user-perceived QoE is defined, based on which the QoS parameters are negotiated and set.

- The QoS parameters are negotiated and set, based on which an assessment of possible QoE metrics is defined.

These protocols can be combined to provide various levels of QoS. The common types of QoS that various vendors may claim to support are as follows:

- **Best Effort QoS:** No QoS is provided.

- **Better Best Effort:** When there is excess bandwidth available after all expedited and assured traffic has been treated, "best effort" traffic is discarded before "better best effort" traffic.

- **Priority-Based QoS:** Superior to best effort because it prioritizes data streams allowing higher priority traffic to be delivered first. If there is too much data of high priority, some data may be lost. High priority data can "starve" lower priority queues.

- **Guaranteed QoS:** Delivers packets according to the specified QoS policy. Can guarantee minimum and maximum bandwidth as well as constant bit rate (CBR) or variable bit rate (VBR) as ATM and frame relay networks have for years

## 2. THE NGN ENVIRONMENT

The issues of NGN environment have been considerably addressed, foremost in ITU-T (ITU-T Rec. Y.2001, 2004; ITU-T Rec. Y.2011, 2004), 3GPP (3GPP TS 23.228, 2006) and ETSI/TISPAN (ETSI ES 282.007, 2006), as well as in recent telecommunications research work. Different logical architectures have been proposed based on the common principles but vary among each other in the logical organization, the services focus and the communications domains.

The generic NGN architecture and its functionalities are represented in Figure 1. A two-layer model is adopted, logically decoupling the transport from the service control functionalities and the services. Four principal groups of functionalities within the NGN architecture can be identified, as follows:

### A   Service-layer application functionalities

The upper-most entities of the service layer represent various general or dedicated application servers (AS), where service logic is hosted and operated. Additionally, the developer-friendly interface functionalities and secure gateway functionalities for third party service provisioning are enabled. The openness and the support for various technologies result in considerable complexity of this NGN segment, and the blended service offering requires mutual engagement and coherent functioning of many application servers simultaneously, therefore orchestration application servers are needed.

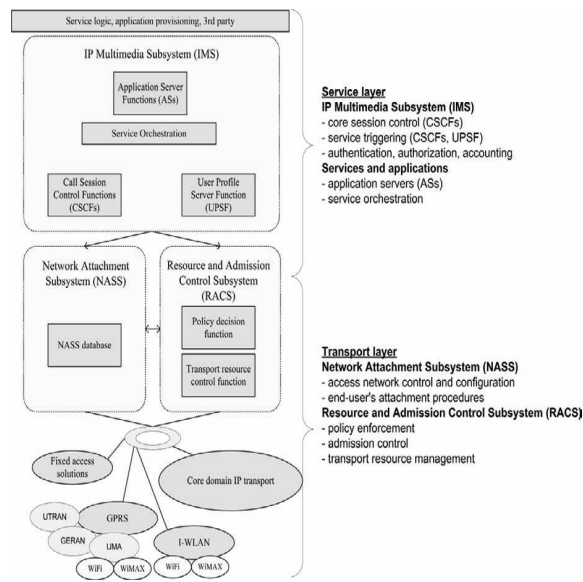

Figure 1. The generic IMS-based NGN model

### B.  Service-layer control segment functionalities:

In this segment, session control, service triggering, and authentication, authorization, and accounting mechanism (AAA) are implemented. Service-layer profiles are sustained here, incoming requests are routed to the appropriate entities and services are triggered. Recently, the IP multimedia subsystem (IMS) (3GPP TS 23.228, 2006; ETSI ES 282.007, 2006) has become the recognized standard for service-layer functionalities and is today incorporated into the majority of recommendations. For this reason, the remainder of this paper assumes the IMS as the core of the service layer. The IMS provides the core session control, service triggering, and authentication and authorization mechanisms for the NGN environment.

### C.  Service-layer to transport-layer arbitrator functionalities

In order to have transparently decoupled service and transport layer, specialized arbitrator functionalities are needed to implement the inter-layer communications and transport control logic. The network attachment subsystem (NASS) is needed that enables the end users admission to the NGN ecosystem and the NGN services, and sustains transport-layer profiles. The resource and admission control subsystem (RACS) performs policy-based resource allocation and appropriate QoS assurance.

*D. Transport-layer functionalities*

The IP based transport platform spans through core and various types of fixed and mobile access networks. It operates under the control of the arbitrator functionalities. The key objective of this group of functionalities is to provide IP connectivity for the purpose of accessing the service-layer functionalities. At this level, the QoS is ensured by using the corresponding mechanisms for the transportation of the media and the reservation, quality, and security accomplishment, which are outside of the scope of NGN. Note that the presented generic NGN model comprises core functionalities that represent the enabling infrastructure for session handling, service triggering, admission control, user management, and quality assurance, whereas additional functionalities are required for specific features, e.g., application-related issues, management, real-time streaming support, access termination, etc.,

## 3. THE IP MULTIMEDIA SUBSYSTEM

The IP multimedia subsystem (IMS), defined by the Third Generation Partnership Project (3GPP) and later adopted by the ETSI TISPAN, has become recognized as the core session control, service triggering, and AAA framework for the delivery of convergent multimedia services within an efficient service delivery environment. Initially it has been proposed as the control subsection of the universal mobile telecommunications services (UMTS) environment, however further expansions have been completed to meet the fixed domain requirements and to address a wider system concept. Nevertheless, both proposals pursue access agnosticism and general user mobility. Logical structuring is clearly defined; session control, user and application data, gateway control and gateways and service environment all reside in clearly separated entities. Interconnection amongst these segments and towards outer world is achieved through open standardized interfaces based on SIP and Diameter protocols and different types of interface technologies.

The basic service provisioning triangle, relevant to this work, consists of the call session control function (CSCF) entities, providing session control, service triggering and AAA functionalities, the home subscriber server (HSS), or the extended user profile server function (UPSF), representing the subscriber profile database and an extended AAA and mobility server, and the application server (AS), hosting the service logic and providing the convergent service delivery environment. Other entities are also defined for the IMS (e.g., media server functionalities, interworking, and gateway functionalities, etc.).

The inherent nature of the IMS as the core session control subsystem is global mobility of end users, services and the ability of these to be independent of the selected access domain and terminal equipment. The IMS-based NGN environment is applicable to both fixed and mobile domains regardless of the initial mobile origin of the IMS subsystem.

However, there are notable mobile characteristics that should be considered that affect the performance of the system as a whole and condition the quality-related issues. For the purpose of quality assurance procedures within the IMS-based NGN environment, the profile entity is important, incorporating relevant subscriber, service and content information. The HSS/UPSF entity of the IMS subsystem sustains the service-layer profile repository, as depicted on Figure 2.

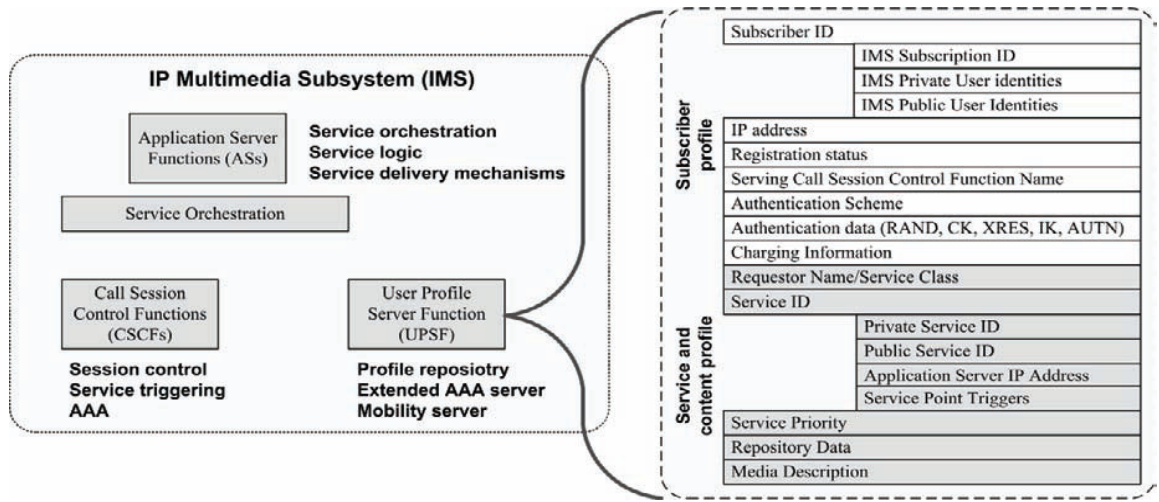

Figure 2. The key quality-related IMS entities and the service-layer profile repository information

## 4. QUALITY ASSURANCE IN THE NGN

### A. *The service-aware Quality assurance approach*

The process of quality assurance in the NGN environment is a challenging task due to several factors. The IP-based next generation environment, originating from Internet domain, is best effort and therefore requires several additional mechanisms to meet the appropriate quality and availability levels. The issue is even intensified due to an extensive range of different media-rich services, which presents a challenge to resource allocation in terms of diverse performance needs (e.g., real-time or near-real-time delivery, priority treatment). In the NGN environment a single session operates across many conceptually and technologically unfamiliar networks, operated by different operators; moreover, the operators do not have full control over the environment as in the legacy telecommunications solutions and each end user is increasingly involved in the shaping of the operation of the environment through the usage of intelligent end user's devices and service personalization.

There are numerous recommendations and guidelines on how to ensure the appropriate IP network level performance objectives (ETSI TS 185.001). However, for complete service delivery, a systematic QoE and QoS assurance is required (ITU-T Rec. Y.1291) that spans through all layers of the solutions and approaches the issue of end user's satisfaction from the services viewpoint rather than from the network viewpoint. Moreover, the notion of multiple separate interconnected domains enforces dynamically changing conditions that imply the usage of dynamic quality assurance mechanisms.

The NGN QoS mechanisms are technology dependent and extend vertically across transport layer and transport control functionalities of the service layer. On the other hand, NGN QoE mechanisms are technology independent and involve service control and application functionalities as well as the mapping of these to transport-layer quality assurance. Only overall integrity and orchestration of all functionalities in all subsystems and layers brings systematic quality assurance in all aspects of service delivery. Based on these prerequisites, the following approach is generally recognized for the NGN environment. The procedure of quality assurance occurs in two stages. First, dynamic negotiation is conducted to set the initial communications parameters in the session set-up procedure. Afterwards, further renegotiations are possible, initiated either by the end user, network, or services.

The QoE and QoS assurance procedures involve vertically the entire NGN environment. On the service layer, the service control and service entities, and profile repositories are engaged, while on the transport layer the user traffic is appropriately handled using various mechanisms (e.g., congestion avoidance, packet marking, queuing and scheduling, traffic classification, policing, and shaping). The resource and admission control entities enforce the arbitrating functionalities that bridge the service and the transport layers. While the entire system is indirectly involved in the QoE and QoS assurance, these functionalities directly enforce the dynamic service-aware admission control and resource reservation, as follows.

### B. *Service-aware iMs-based ngn Quality assurance Procedure*

The resource and admission control subsystem (RACS) has become generally recognized as the subsection of the NGN responsible for the policy control, resource reservation and admission control. Standardization efforts of ETSI TISPAN NGN and ITU-T have addressed the issue of policy-based admittance of the end user to the resources based on a rather complex service-aware procedure of negotiation. The proposals vary in the defined entities and logical organization but are conceptually similar and extend horizontally cross access and core domain and vertically across service and transport layers. As depicted in Figure 3, the generic RACS comprises:

- The policy decision function, negotiating with the session control and application functions via northbound interfaces.

- The transport resource control functions, representing the mediator between the policy decision function and the transport infrastructure through dedicated permission control mechanisms.

- The transport policy enforcement functions, residing on the transport infrastructure and enforcing the final quality-related decisions.

The policy decision function represents the mediation layer between the service provisioning domain and the network resource-provisioning domain, providing an appropriate level of abstraction of the resource processing technologies to the service execution technologies. The policy decision function issues a request for resource authorization and reservation, indicating the QoS characteristics (negotiated with the service provisioning domain). The resource control function is in charge of the permission control mechanisms and informs the policy decision function of the successful resource allocation.

In general, separate resource control functions exist for the core network and for each type of access network, taking into account specific characteristics and management policy. In the process of the resource allocation it consults the network attachment subsystem (NASS) for the access and transport-layer QoS profile. Other functionalities of the RACS are the border gateway functions and the resource control enforcement functions that perform the gate control, packet marking, resource allocation, network address translation, policing and usage metering, etc. In general, the resource control functions act as the local policy decision points in terms of subscriber access admission control and resource handling control, whereas the policy decision function represents the final policy decision point.

The resource control function derives and installs the Layer 3 and Layer 2 traffic policy, indicating the traffic control handling (e.g., gate control, packet marking, etc.). In the process of granting the resources the network QoS parameters of the Layer 3 and Layer 2 are mapped to the respective policy. The operation of the RACS is generally application agnostic but supports traffic control for the purpose of application delivery with Uni-/bidirectional, a-/symmetric, Uni-/multicast, up-/downstream traffic patterns.

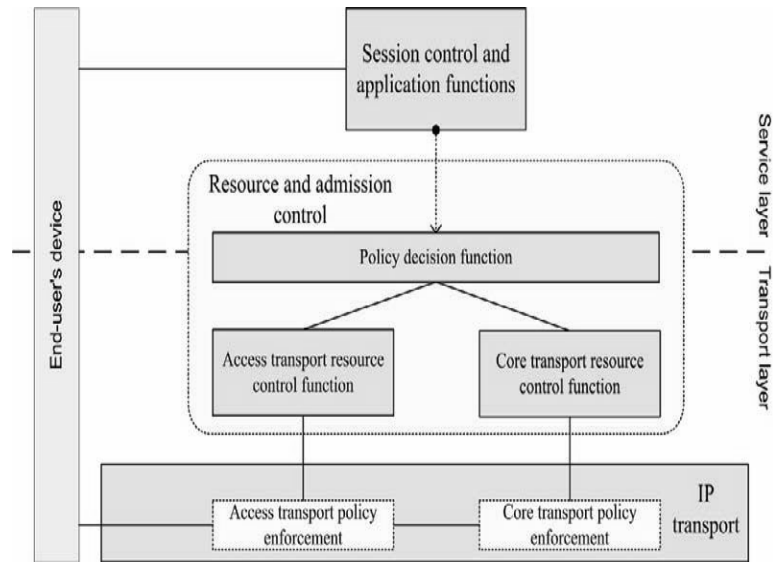

Figure 3. Resource and admission control subsystem (RACS)

## C. The network attachment

The network attachment subsystem (NASS) has also been considered for the NGN-based environment within the standardization efforts of ETSI TISPAN NGN and ITU for the purpose of consistent and controlled registration and attachment of the end users accessing the NGN services through various access networks. The NASS is responsible for the registration procedures within the access domain and the initialization of the end user's terminal equipment when accessing the

## D. 3GPP End-to-End QoS framework

The term "Quality of Service" sums up all quality features of a communication as perceived by a user for a specific service. In order to achieve the end-to-end QoS, it is necessary to maintain a level of QoS all along the path from the source TE (Terminal Equipment) to the destination TE crossing various administrative domains. In the context of IMS services, the involved domains will be NGN Bearer Service domain, external IP domain, IMS domain and/or other UMTS Bearer Service domains. The 3GPP proposes the use of Diffserv to support QoS in the underlying IP networks. Furthermore, the provisioning of QoS is performed by the PBM framework standardized by the IETF [11, 12, 13].

DiffServ provides a scalable aggregate approach to categorize into different classes that are subjected to a specific treatment, known as PHB (Per Hop Behavior). IETF defines three main groups of classes: EF (Expedited Forwarding), AF (Assured Forwarding) and BE (Best Effort). The EF class aims to provide low loss, low delay and low jitter guaranteed services. The AF class gives different forwarding assurances in terms of loss, delay and jitter. It is composed of a set of Policy Enforcement Point (PEP), a Policy Decision Point (PDP) and a Policy Repository component. The PEP component is a policy decision enforcer located in the network and system equipments. The PDP is a decision-making component that governs the logic of the overall management system based on the high level directives of the administrator/operator based on the agreed SLA (Service Level Agreement) with his customers.

A good QoS system supports standards so that each network component interacts in a heterogeneous networking environment comprised of different vendor's equipment. As a network administrator you may not always be in control of the type of equipment that will be included in your network. As a result of acquisitions, you may find yourself faced with an International journal on applications of graph theory in wireless ad hoc networks and sensor networks (GRAPH-HOC) Vol.2, No.2, June 2010integration scenario that will be much easier to address if your networking equipment supports standard protocols that allow QoS functionality to be mapped between the various layers resulting in effective, heterogeneous networking.

## 5. NGN SERVICES, PROVIDING IDENTIFICATION AND AUTHENTICATION

On the network level, management of IP addressing scheme within the access networks and authentication of the access sessions. The following key functionalities are provided through the NASS:

- Dynamic allocation of IP addresses and other relevant parameters for the end user's terminal equipment configuration
- IP-layer authentication before or within the procedure of IP address allocation
- Network access authorization based on the subscriber profile
- Access network configuration based on the subscriber profile
- IP-layer location management

| | | Globally Unique IP Address/Realm | Logical Access ID |
|---|---|---|---|
| TRANSPORT-LAYER ACCESS, SESSION AND SUBSCRIPTION INFORMATION | ACCESS SESSION DECRIPTION | Subscriber ID | Access Networky Type |
| | | Physical Access ID | RACS Point of Contact |
| | | Privacy Indicator | |
| | | Terminal Profile Hardware (e.g., model, display size, resolution, processor & memory info, sound capabilities) Network Connectivity (e.g., supported interfaces, current interface, DL and UL capabilities) Software (e.g., OS, browser info, supported media types, supported content protection) User Preferences (e.g., desired/acceptable service quality, time&budget constraints) | |
| | QoS profile | Transport Service Class | Requestor Name |
| | | Maximum Priority | Media Type |
| | | UL Subscribed bandwidth | DL Subscribed bandwidth |
| Initial Gate Settings | | List of allowed destinations | UL Default bandwidth |
| | | DL Default bandwidth | |
| Physical and logical access ID | | Location Information | Derault Subscriber ID |
| | | RACS Point of Contact | Access Network Type |

Figure 4. Transport-layer access, session, and subscription information

The functionalities are provided through several logical entities. Among these, the functionality responsible for session description and transport layer profile maintenance is actively involved in the quality assurance procedure (referred to as the NASS database—NASS DB). It communicates with the RACS subsystem to relay the relevant transport-layer access, session and subscription information, involved in the quality assurance procedures. An example of the information model of the NASS is represented in Figure 4.

### A. Service-aware IMS-based NGN Quality Assurance Procedure

Referring to Figure 5, within the generic session set-up procedure, the following steps are involved in complete NGN QoE and QoS assurance:

- Service authentication procedure based on the requesting user and the requested service
- Parameter negotiation and resource authentication
- Determination of final feasible service configuration and final application operation point based on resource allocation capabilities

68

- Final profile confirmation and delivery of the requested service to the end user

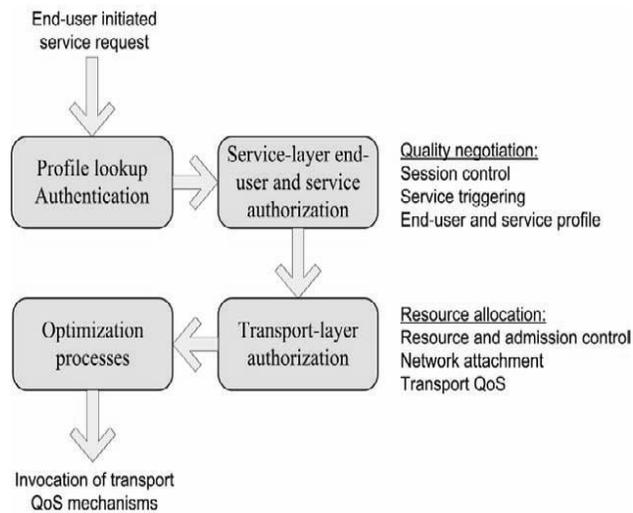

Figure 5. The generic NGN quality assurance procedure

According to ITU-T Rec. Y.2111, there are three basic scenarios for QoS provisioning, as follows:

1. The end user is only aware of the services they may request and is unaware of the QoS signaling mechanisms, while it is the responsibility of the service control functionalities to determine the QoS service requirements and issue the respective requests to resource authorization functionalities. The latter perform resource authorization and reservation procedures.

2. The end user's device is capable of signaling and managing its QoS resources, however prior authorization via the service control functionalities is required. After the initial service request, the service control functionality determines the QoS service requirements and requests the network authorization. If approved, the end user's device receives the authorization token and requests resource reservation.

3. The end user's terminal is capable of issuing QoS requests over signaling and management protocols without prior authorization. These scenarios are heavily dependent of the operator's policy standpoint, defining the strictness of the resource allocation and the service awareness of the transport-layer operation. The notion of heterogeneous access domain, user and service mobility, and support of various end users devices in a multimedia-oriented NGN environment requires a generalized approach that assumes any of the above scenarios. From the end user's terminal devices viewpoint the first scenario is most general, while the remaining two scenarios could be understood as simplified cases of the first scenario. Therefore, any further discussions are in terms of the first scenario.

Based on the IMS-based NGN architecture presented before and the dynamic service-aware approach to quality assurance for the NGN as the following:

1. The quality assurance procedure consists of two consecutive sections. The first section involves service-layer IMS-based service provisioning and user authentication and authorization procedures. Based on the service request, issued by the end user's device, *the IMS CSCF entities perform authentication and authorization procedures based on the information, retrieved from the UPSF.*

2. *If successful, the CSCF entities issue a request for resource authorization and reservation to the RACS policy decision function, containing the parameters of the requested transport control service (i.e., priority and QoS parameters).*

3. *The second section involves transport-layer RACS based service policy definition and resource allocation procedures. The policy decision function receives the request, chooses the service policy and performs an authentication procedure.*

4. *The authentication procedure is based on the process of matching the requested parameters against the chosen service policy. If successful, the policy decision function is in charge of forwarding the request to the resource control functions that have been chosen through the service policy.*

## 6. CONCLUSION

Within the NGN, the policy-based quality assurance QoS and QoE seems to be the reasonable approach. The characteristics of the NGN environment present challenges to quality assurance for several reasons, such as general support of mobility, access agnosticism, multi-domain environment, best-effort technologies, etc. While mechanisms and technologies for the transport-layer core and access quality assurance are well defined, the issues of interconnection and interworking need to be resolved in order to achieve dynamic service-aware end-to-end user-perceived quality experience. The proposal presented here is an approach that engages all layers of the environment via parameterization, profiling, negotiation, and arbitrating mechanisms, pursuing the end-to-end controllability of the quality of the respective communications. QoS is a vital component of any network. QoS is even more critical for converged networks. As soon as your network is required to support traffic that is sensitive to delay or packet loss, QoS must be present to provide the assurances that these data flows are delivered with timeliness without dropping packets. Adding bandwidth to your network might appear to be a cheaper solution, but the unpredictable nature of network traffic flows can result in momentary congestion. If QoS is not present, VoIP and video traffic will suffer from excessive delay and packet loss rendering them ineffective.

Further challenges arise from this concept. The complexity and the performance requirements of the rather complex signaling procedures are an issue that would present a substantial load to the entire environment, and the required level of intelligence needed to perform the quality negotiation and enforcement with the respective security issues is challenging. Further standardization efforts would be required to resolve the interconnection and interworking of the traversed access and transport domains as well as with the service-layer mechanisms. Once the various proposals are harmonized and the standardization is completed, the NGN services can be considered as a collection of specialized services implemented with the already available functionalities of the NGN environment in a standardized fashion and with ensured operator-grade quality.

## REFERENCES

[1] 3GPP TR 23.802 (2005). Technical Specification Group Services and System Aspects – Architectural enhancements for end-to-end Quality of Service (QoS).

[2] 3GPP TS 23.228 (2006). Technical Specification Group Services and System Aspects: IP Multimedia Subsystem (IMS), Rel. 7.

[3] Ban, S. Y., Choi, J. K., & Kim, H. S. (2006). Efficient end-to-end qos mechanism using egress node resource prediction in NGN network. ICACT 2006, 1, pp. 480-483.

[4] Cicconetti, C., Lenzini, L., & Mingozzi, E. (2006). Quality of service support in IEEE 802.16 Networks. IEEE Network, 20(2), 50-55.

[5] Dixit, S., Guo, Y., & Antoniou, Z. (2001). Resource management and quality of service in third-generation wireless networks. IEEE Commun. Mag., 39(2), 125-133.

[6] DSL Forum TR-126. (2006). Triple-play Services Quality of Experience (QoE) Requirements.

[7] ITU-T Rec. Y.2111 (2006). Next Generation Networks – Quality of Service and Performance: Resource and admission control functions in Next generation Networks.

[8] Kapov, L. S., & Matjasevic, M. (2006). End-to-end QoS signaling for future multimedia services in the NGN. LNCS, Vol. 4003. Springer.

[9] Park, S. et al. (2003). Collaborative QoS architecture between DiffServ and 802.11e wireless LAN. VTC 2003, 2, pp. 945-949.